\documentclass[aps,prl,superscriptaddress,twocolumn,showpacs,10pt]{revtex4-2}

\usepackage{graphicx}
\usepackage[ansinew]{inputenc}
\usepackage[normalem]{ulem}
\usepackage{color}
\usepackage[usenames,dvipsnames]{xcolor}
\usepackage{multirow}
\usepackage{colortbl}
\usepackage{amssymb}
\usepackage{soul}

\newcommand{\ieap}{Institut f\"ur Experimentelle und Angewandte Physik, Christian-Albrechts-Universit\"at zu Kiel, D-24098 Kiel, Germany}
\newcommand{\thh}{Institut f\"ur Theoretische Physik, Universit\"at Hamburg, D-20355 Hamburg, Germany} 
\newcommand{\jul}{Peter Gr\"{u}nberg Institut and Institute for Advanced Simulation, Forschungszentrum J\"{u}lich \& JARA, 52425 J\"{u}lich, Germany}
\newcommand{\due}{Faculty of Physics, University of Duisburg-Essen and CENIDE, 47053 Duisburg, Germany}
\newcommand{\supercomp}{J\"ulich Supercomputing Centre, Forschungszentrum J\"{u}lich and JARA, 52425 J\"{u}lich, Germany}

\newcommand{\sm}{Supplemental Material}

\newcommand{\didv}{$dI/dV$}

\begin{document}

\title{Spin excitation of Co atoms at monatomic Cu chains}

\author{Neda Noei} \affiliation{\ieap}
\author{Roberto Mozara} \affiliation{\thh}
\author{Ana M. Montero} \affiliation{\jul}
\author{Sascha Brinker} \affiliation{\jul}
\author{Niklas Ide} \affiliation{\ieap}
\author{Filipe S. M. Guimar\~aes} \affiliation{\supercomp}
\author{Alexander I. Lichtenstein} \affiliation{\thh} 
\author{Richard Berndt} \affiliation{\ieap}
\author{Samir Lounis} \affiliation{\jul} \affiliation{\due}
\author{Alexander Weismann} \email{weismann@physik.uni-kiel.de} \affiliation{\ieap}

\begin{abstract}
The zero-bias anomaly in conductance spectra of single Co atoms on Cu(111) observed at $\approx 4$~K, which has been interpreted as being due to a Kondo resonance, is strongly modified when the Co atoms are attached to monatomic Cu chains. 
Scanning tunneling spectra measured at 340~mK in magnetic fields exhibit all characteristics of spin-flip excitations.
Their dependence on the  magnetic field reveals a magnetic anisotropy and suggests a non-collinear spin state. 
This indicates that spin-orbit coupling (SOC), which has so far been neglected in theoretical studies of Co/Cu(111), has to be taken into account.
According to our density functional theory and multi-orbital quantum Monte Carlo calculations SOC suppresses the Kondo effect for all studied geometries.
\end{abstract}

\maketitle

Atoms with localized unpaired spins coupled to a non-magnetic metal can exhibit the many-body Kondo effect, which was first observed from metals with magnetic impurities via a resistivity increase at low temperatures \cite{de_haas_electrical_1934}.
J. Kondo explained this effect in terms of exchange scattering processes that flip the localized spin and the spin of the scattered conduction electron \cite{kondo_1964, kondo_1968}.
Below a characteristic Kondo temperature $T_K$ the magnetic moment of the impurity is screened by conduction electrons leading to a singlet ground state.
$T_K$ also determines the width of the Kondo resonance that develops in the single-particle spectrum of the impurity around the Fermi energy $E_F$.
Developing theoretical techniques that properly describe the impurity spectrum and the thermodynamic properties of Kondo systems remained a challenge for decades \cite{hewson}.
Spectroscopy of single magnetic impurities with the scanning tunneling microscope (STM) revealed an anti-resonance in the differential conductance of Ce/Ag(111) \cite{li_1998, ter09} and Co/Au(111) \cite{madhavan_1998} that was attributed to a Fano interference \cite{fano_1961} between tunneling processes into localized $f$- or $d$-orbitals and the conduction band. 
Subsequent studies investigated transition metal atoms on different substrates and in modified environments \cite{jam01, mad02, nag02, schn02, otte_role_2008, liq18, bork_2011, moro_2018}, in particular, Co at Cu surfaces \cite{wahl98, mano00, knorr_2002, wahl_2004, neel_2007,  vita08,  neel_2008, neel_2010, neel_2011, Neel_2019, figg19, Neel_2020}.
  
As to Co/Cu(111), some experimental observations deviate from theoretical expectations for Kondo systems. 
First, the Kondo resonance line shape is not Lorentzian and therefore Fano resonances should not match the experimental spectra well. 
Rather, a Frota line shape is expected close to $E_F$ and a logarithmic decay at larger absolute energies \cite{frota_1992}. 
This is because Kondo systems are local Fermi liquids \cite{nozieres_1974} for which the self energy $\Sigma(E)$ close to $E_F$ and consequently the life-time broadening depend on the electron energy $E$ \cite{Bulla_1998} whereas an energy independent $\Sigma$ leads to a Lorentzian line-shape (see \sm~for details).
Frota line shapes indeed agree better with experimental data for Co and Ti on Cu$_2$N  \cite{von_bergmann_spin_2015, Osolin_2013,Zitko_2011}, for molecular Kondo systems \cite{karan_2015, ormaza_2017, gruber_2018, karan_2016}, and for subsurface Fe in Cu(100) \cite{pruser_2011, pruser_2012}. 
For Co on noble metal surfaces, however, Fano lines are superior \cite{Neel_2020, Neel_2019}. 
Second, the experimental resonance consistently occurs above $E_F$ for Co on Cu(111), Ag(111) and Au(111) \cite{knorr_2002, wahl_2004, moro_2018, bork_2011} whereas the Friedel sum rule predicts a Kondo resonance below $E_F$, because the Co $d$-shell is more than half filled \cite{hewson}. 
Theoretical studies found the resonance at $E_F$ when modeling Co as a spin-1/2 system \cite{Aligia_2021}, and below $E_F$ when the multi-orbital nature of the $d$-shell is taken into account \cite{surer_2012, Neel_2020}. 
Modeling of Co/Ag(111) by a single electron residing in two orbitals led to a SU(4)-Kondo resonance above $E_F$ \cite{moro_2018}.
To match experimental data, however, the calculated spectrum was shifted to positive voltages by 4~meV\@. 

Recently, a different explanation of the zero-bias feature has been presented, namely a spin-polarized quasiparticle denoted spinaron \cite{bouaziz_2020}.
It reflects a renormalized electronic density of states caused by inelastic spin excitations (ISE) and develops above $E_F$ for the case of Co atoms on Cu, Ag and Au(111) surfaces.
The energy of ISE is non-zero without a magnetic field due to magnetic anisotropy (MA).
Generally, the importance of MA for transition metal atoms on decoupling layers is well established \cite{hirjibehedin_2006, hirjibehedin_2007, otte_2009, otte_2008} and inelastic tunneling spectra calculated from quantum spin models reproduce experimental data \cite{ternes_2015}. 
Spin excitations have also been reported from of Fe adatoms on metals \cite{khajetoorians_2011}.		

Discriminating between ISE and Kondo resonances can be difficult.
Kondo resonances may occur below or above $E_F$ depending on the $d$-orbital filling \cite{hewson, wahl_2004} and Fano-like interference can render their line shapes asymmetric. 
On the other hand, ISE spectra are to lowest order composed of broadened steps located symmetrically around zero bias at $eV=\pm \Delta_i$, where $\Delta_i$ is a spin excitation energy \cite{ternes_2015}.
Asymmetry may result from a coherent superposition of potential and exchange scattering  \cite{FePorphy_2018,ternes_2015,Schweflinghaus_2014}, a spin-polarization of the tip \cite{Loth_Nat_2010,Loth_NJP_2010}, and the aforementioned spinaronic state  \cite{bouaziz_2020}. 
The width of the inelastic steps is affected by temperature and by coupling to electron-hole excitations of the substrate that reduce the lifetime of the excited spin state \cite{muniz_2003, Lounis_2010,khajetoorians_2011}.
Higher excitation energies $\Delta_i$ and stronger coupling to a substrate thus lead to broadening, which may be represented by a higher effective temperature within the framework of inelastic tunneling spectroscopy (IETS). 

The large width of the anti-resonance observed from Co adatoms on the Cu(111) surface implies that a direct experimental test of the role of MA would require extremely large magnetic fields. 
Here, we bring Co atoms on Cu(111) close to monatomic Cu chains.
The increased coordination is expected to enhance the electron density available for hybridization, which would strengthen the Kondo effect if already present for the isolated adatom.  
The breaking of the approximate rotational symmetry around the surface normal reduces the angular momentum and hence the MA as desired.
Nevertheless, our measurements in a magnetic vector field at sub-K temperatures clearly favor ISE over a Kondo effect.

Experiments were performed with a STM in ultrahigh vacuum at 340~mK with a magnetic field up to 9~T (2~T) perpendicular (parallel) to the surface (Unisoku USM1300).
To prevent diffusion Co was deposited onto the cold sample in the STM from a custom-built electron beam evaporator.
Monatomic Cu chains were prepared as described in Ref.~\cite{noei_apparent_2018}.

\begin{figure}
\centering \includegraphics[width=\linewidth]{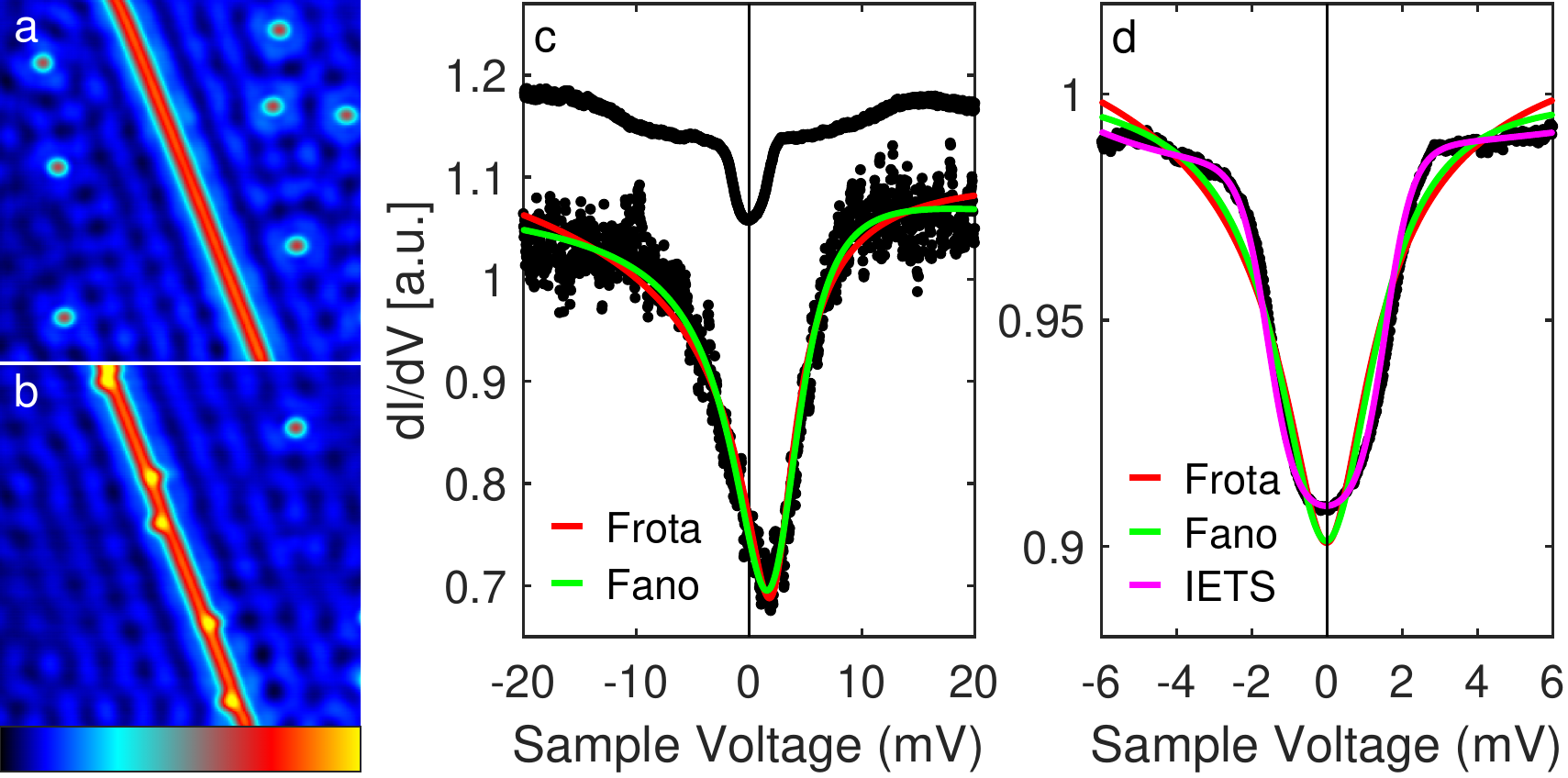}
\caption{
(a) STM image ($V=30$~mV, $I=30$~pA) of Co atoms (red dots) and a monoatomic Cu chain (red stripe) on Cu(111).
Blue undulations are due to scattering of surface state electrons.
(b) The same area as in (a) imaged after moving six Co atoms (yellow protrusions) to the chain with the STM tip.
(c) \didv\ spectra of Co atoms on the (111) surface (lower data) and at a Cu chain (top, shifted upward by 0.15) normalized to a reference \didv spectrum taken off the atom.
Fitted Fano and Frota lines are shown by green and red lines, respectively.
(d) Magnified view of the spectrum from Co at the chain.
Best Fano and Frota fits systematically deviate from the data while a model function for spin excitations (magenta) exhibits a good match.}
\label{topo}
\end{figure}

Figure~\ref{topo}a displays a constant current topograph with seven Co atoms (red dots) six of which were subsequently moved to the sides of a monatomic Cu chain (Fig.~\ref{topo}b, yellow protrusions).
$dI/dV$ spectra (Fig.~\ref{topo}c) were recorded above the Co atoms before and after attachment to the chain.
The Co atoms on the pristine surface display an antiresonance similar to previous data \cite{knorr_2002, Neel_2020, Neel_2019} with a reduction of the differential conductance by $\approx 30$\% centered around $V\approx 3$~mV
\footnote{We verified that a possible experimental voltage offset was less than 50~$\mu$V\@.}. 
The Co atoms at the chain show a smaller anti-resonance (reduction $\approx$ 10\%) that is also narrower and centered around $V=0$.
The spectral shape is remarkably similar to the steps observed from systems with inelastic spin excitations. 
Additional broader features appear at $V\approx \pm 12$~mV\@. 

\begin{figure*}
\centering \includegraphics[width=0.9\linewidth]{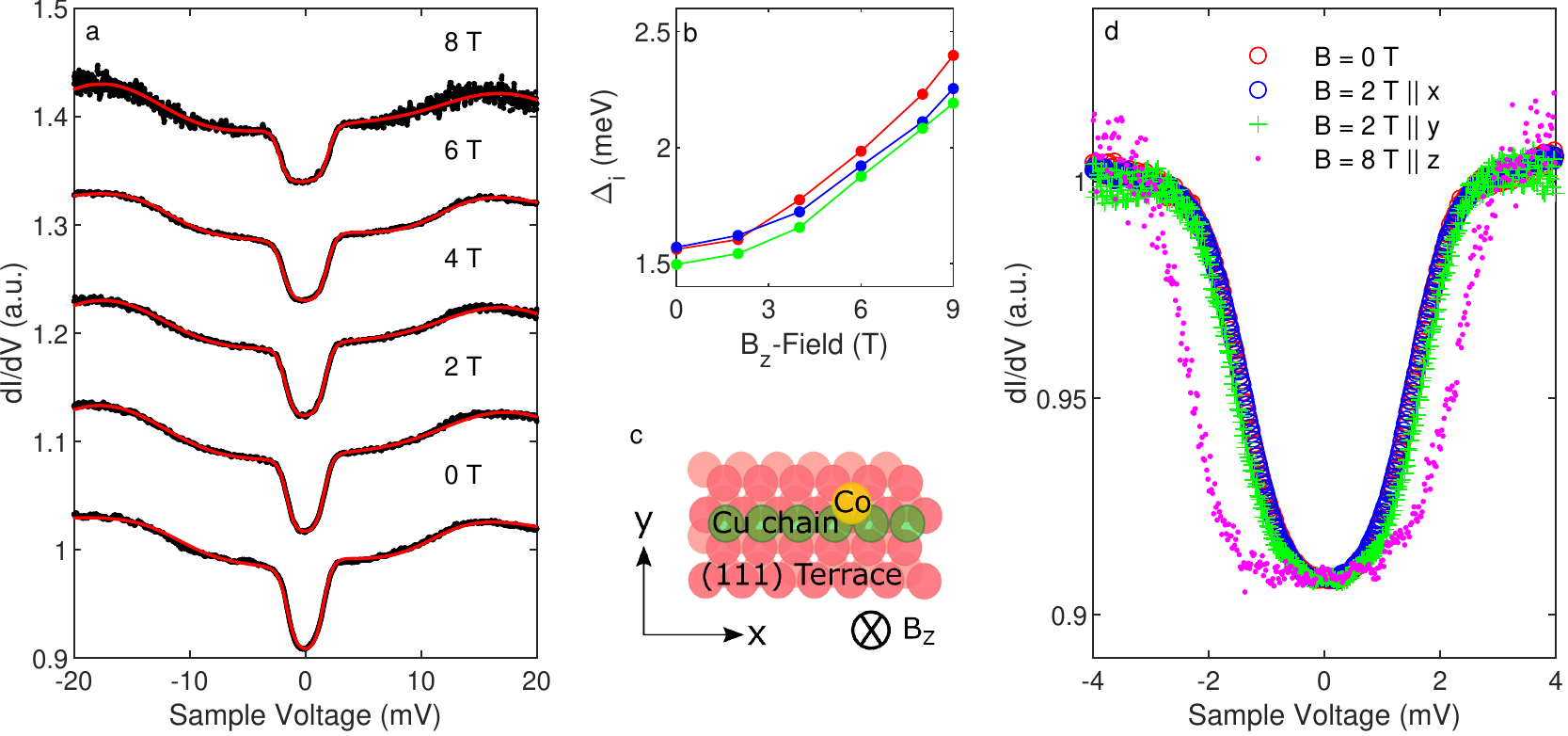}
\caption{\textbf{(a)} Evolution of \didv\ spectra of a Co atom at a Cu chain with the strength of an out-of-plane magnetic field, $B_{z}$.
Spectra were divided by reference spectra taken on the chain away from the atom to display relative conductances. The individual spectra were shifted vertically by 0.1 for increasing B-field; 
\textbf{(b)} Excitation energy $\Delta_i$ of the steps near $\pm2$~mV vs.\ $B_z$ recorded above three Co atoms at chains as obtained from an IETS fit.
Despite some scatter all data sets display a non-linear increase of $\Delta$.
\textbf{(c)} Sketch of the coordinate system defining the in-plane field directions.
\textbf{(d)} $dI/dV$ spectra for different field directions. For better comparison, spectra were rescaled and shifted individually so that upper and lower conductance plateaus overlap. The spectrum measured in a field of 2~T parallel to the chain hardly deviates from the field-free data. In contrast, for $B_y$=2~T an $\approx 19$~\% increase of $\Delta$ is observed.}
\label{Bz}
\end{figure*}

Fits of the experimental data are compared in Figs.~\ref{topo}c and d.
For an isolated atom, the Fano resonance agrees better than the Frota line shape.
However, neither Fano (green) nor Frota (red) lines match the data from Co at a Cu chain well whereas good agreement is achieved with a model of two IETS steps \cite{ternes_2015} at $\approx2$ and 12~meV with different effective temperatures (Figs.~\ref{topo}d, magenta line, and \ref{Bz}a). 

Next, the effect of a magnetic field $B$ perpendicular to the crystal surface ($z$-axis) on the line shape is explored (Fig.~\ref{Bz}).
The spectra reveal an upward shift of the low-energy step as $B$ increases. 
The increase is nonlinear indicating that the spin quantization axis is not aligned with the magnetic field direction.
A non-zero excitation energy without magnetic field directly indicates a substantial MA\@. 
The zero-field width of the step near 2~meV corresponds to an effective temperature $T_{\text{eff}} = 3.2 \pm 0.2$~K while the $\pm 12$~mV step is broader ($T_{\text{eff}} = 20 \pm 2$~K) and consequently no $B$ dependence is resolved. 

The field strengths available parallel to the crystal surface were limited to 2~T\@.
Parallel to the chain (Fig.~\ref{Bz}d), $B_x$), no significant change of the  excitation energy is observed.
In the perpendicular direction ($B_y$), $\Delta$ increases by $\approx13$\%, similar to the change of $\Delta$ at $B_z=2$~T\@.

The data demonstrate that the spectral feature observed from Co at chains is due to inelastic spin excitations.
The underlying MA essentially confines the magnetic moment to the $yz$ plane.
Since MA results from spin orbit coupling (SOC) and crystal field splitting of the $d$-levels \cite{dai_2008}, our observations require that SOC must not be neglected in modeling the spectra of Co/Cu(111).

\begin{figure}
	\centering
		\includegraphics[width=0.9\linewidth]{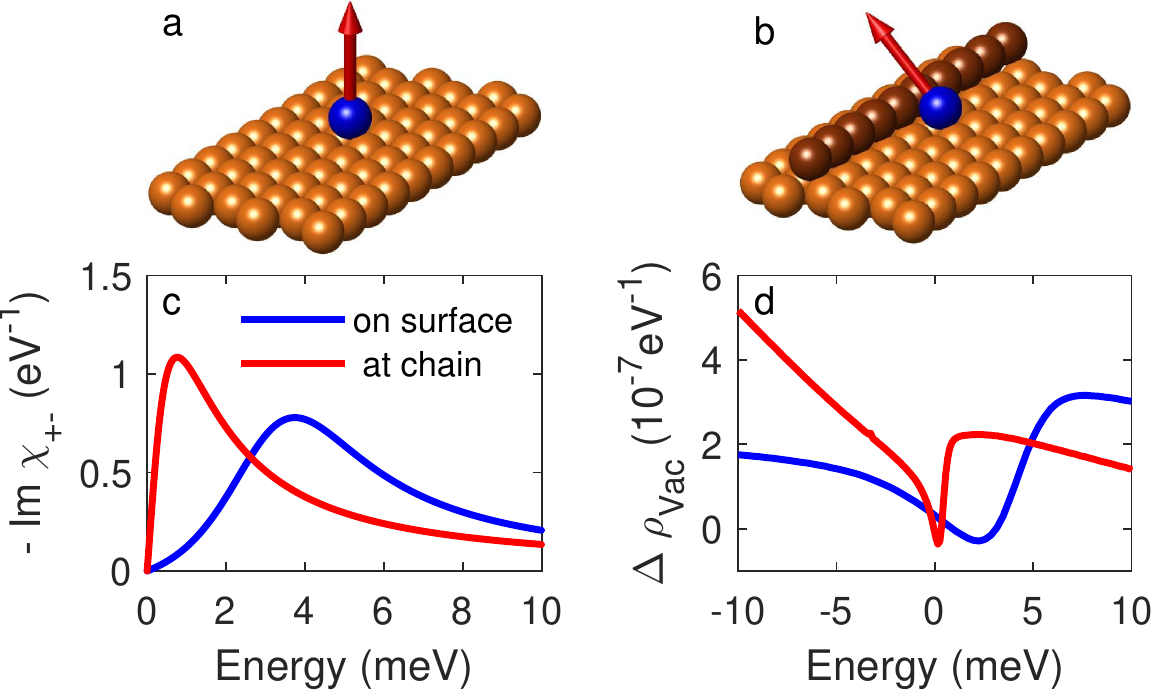}
		\caption{Magnetic states  of the isolated Co adatom on (a) Cu(111) and (b) attached to the Cu chain.
		(c) Imaginary part of the dynamical transverse spin susceptibility from TD-DFT defining the spectral density of spin-flip excitations. 
		(d) Theoretical inelastic STS spectrum corresponding to the local density of states in vacuum as renormalized from MBPT, calculated at a height of 470~pm above the adatom. }
	\label{fig:MBPT}
\end{figure}

The MA of Co adatoms was calculated with density functional theory (DFT) including SOC via the Korringa-Kohn-Rostoker Green function (KKR) method~\cite{Bauer_2014}. 
For Co on the flat Cu surface we find a $d$-occupation $n_d=7.45$ electrons, and spin and orbital moments $m_s=1.99\mu_B$ and $m_L=0.44\mu_B$. 
The MA is uniaxial with the easy axis oriented perpendicular to the surface.
The anisotropy barrier for rotation of the magnetic moment into the $xy$-plane is 1.78~meV\@.
Attaching Co to a Cu chain hardly alters the magnetic moments ($m_s=1.88$, $m_L=0.49$ in Bohr magnetons) or the occupation ($n_d=7.48$) but tilts the spin.
As illustrated in Fig.~\ref{fig:MBPT}~b, the calculated easy axis is rotated by $\approx45^{\circ}$ from the $z$ direction towards the chain (Fig.~2 of the \sm), \emph{i.\,e.} the moment lies in the $yz$ plane and the hard axis is parallel to the chain.
These results match the experimentally observed dependencies on the magnetic field vector.
DFT therefore suggests a non-collinear spin-state to be present at the side of the chain. 

Magnetic anisotropy being relevant for the spectra at the side of the chain, it should also be pivotal for isolated Co adatoms, which experience larger MA according to DFT\@.
MA affects the Kondo behavior in a non-trivial fashion.
To obtain a Kondo effect, two degenerate spin states must be lowest in energy and connected by a single spin-flip process with $\Delta m_S=\pm1$ \cite{otte_2008}. 
Kondo screening may persist if these levels are split by less than $k_B T_K$ \cite{jacob_2018}. 
For $S=3/2$ with easy plane MA, which is the case of Co on Cu$_2$N, a Kondo resonance remains intact even at MA values exceeding $k_B T_K$, because the ground states with $m_S=\pm1/2$ remain degenerate.
However, in the case $S=3/2$ with easy axis MA, and in all $S=1$ scenarios, a sufficiently large MA quenches the Kondo effect or reduces $T_K$ exponentially \cite{Zitko2008, Blesio2019}. 
In other words, the Kondo effect is unlikely to be present for Co adatoms that exhibit an out-of-plane easy axis as described above.

We used two approaches to model the conductance spectra.
The first one is based on time-dependent density functional theory (TD-DFT)~\cite{Lounis_2010,Dias_2015} coupled to many body perturbation theory (MBPT)~\cite{Schweflinghaus_2014} and accounting for SOC~\cite{bouaziz_2020}\@. 
It reproduces the zero-bias anomalies of Co adatoms on Cu(111) and other surfaces \cite{bouaziz_2020}.
The second approach is the multi-orbital continuous-time quantum Monte Carlo (QMC) method~\cite{gull_2011} that more realistically captures electron correlations including the Kondo effect and was previously applied for modeling Co/Cu(111)~\cite{surer_2012}. 
Here, we extend this investigation to include SOC\@. 

The procedure described in Ref.~\cite{bouaziz_2020} consists in extracting first the dynamic transverse spin-susceptibility, $\chi$, whose imaginary part gives the spectral density of spin-flip excitations (Fig.~\ref{fig:MBPT}c).
It contains two essential properties: (i) a peak at the effective anisotropy barrier and (ii) a broadening directly related to Stoner electron-hole excitations induced by electrons of the metallic substrate.
The attachment of the Co atom to the Cu chain reduces the MA and therefore the peak energy. 
The interaction between electrons and the intrinsic spin-excitation is quantified within MBPT and renormalizes the theoretical STS spectra obtained within the Tersoff-Hamann approximation~\cite{Tersoff_1985}. 
Like in the experimental data, the spectral feature at the chain (Fig.~\ref{fig:MBPT}d, red) is narrower and shifted toward $E_F$.
For the isolated adatom, the antiresonance is recovered
\footnote{As detailed in the \sm, a fit of the theoretical spectra is better reproduced with Fano than with Frota lineshapes in line with the experimental data.}.
It results from a conventional spin-excitation giving rise to a conductance step at negative bias and overlaps with a step at positive bias induced by a spinaron, a bound state emerging from the coupling of electrons and spin-excitations \cite{bouaziz_2020}. 
When the atom is surrounded by the wire atoms, the two steps acquire similar heights with a reduced gap as expected from the weaker anisotropy barrier. 
We note that MBPT is so far not able to account for non-collinear magnetic states. 
Therefore, the spectra for both geometries were calculated assuming an out-of-plane orientation of the magnetic moment.

For the QMC calculations we determined the crystal field splitting of the $d$-orbitals and orbital dependent hybridization functions with DFT without SOC and then added SOC and the multi-orbital Coulomb interaction $U$ to the ionic Hamiltonian before performing a hybridization expansion within QMC to calculate the self-energy $\Sigma(i\omega_n)$.
We verified that Hund's rules are obeyed when hybridization to the substrate and crystal field splitting are switched off. 
Calculations were done for a $d^8$ configuration in agreement with previous work \cite{surer_2012} and the magnetic spin moment from the aforementioned DFT calculations.
QMC is performed in imaginary time interval $\tau=0 \dots (k_BT)^{-1}$ and thus the computational effort grows when the temperature is lowered. 
Since we used $(k_BT)^{-1}=200$~eV$^{-1}$ the corresponding spectral broadening precludes a quantitative description of sub-meV details of the STS spectra.

\begin{figure}
	\centering
		\includegraphics[width=0.9\linewidth]{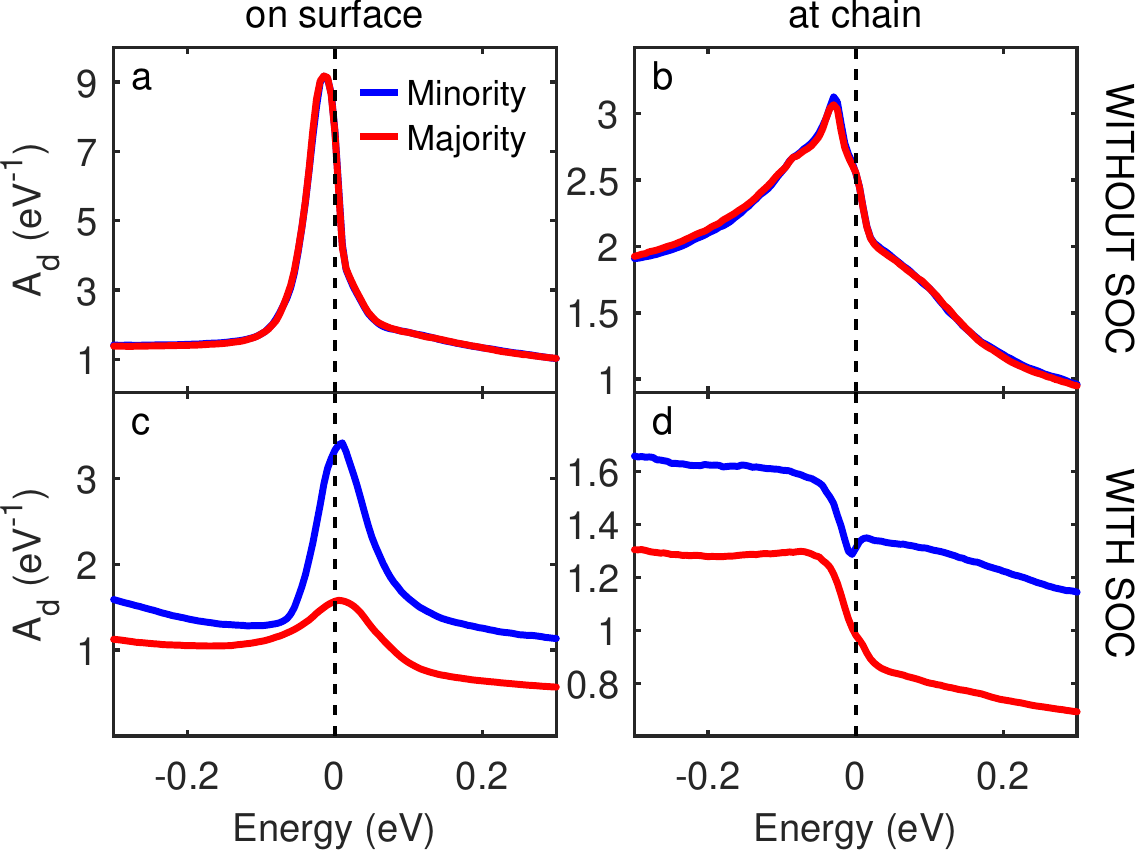}
		\caption{Low energy part of the spin resolved spectral functions $A_d(\omega)$ of the Co \textit{d}-states from QMC without (a, b) and with (c, d) SOC for the free Co atom (a, c) and the Co atom attached to the side of the chain (b, d), respectively. Spectral densities for majority (minority) spin direction are plotted in blue (red).
		Without SOC, both are identical within numerical uncertainties. With SOC, a magnetic ground state develops and a spin-polarization is present.}
	\label{fig:QMC}
\end{figure}

Figure \ref{fig:QMC} displays the total \textit{d}-orbital spectral functions close to $E_F$.
Orbital resolved results for a larger energy range may be found in the \sm. 
For Co on the pristine surface, a resonance below $E_F$ is observed without SOC (Fig.~\ref{fig:QMC}a). 
For Co at a Cu chain, this resonance becomes weaker, broader, and more asymmetric (Fig.~\ref{fig:QMC}b).
In both geometries, the resonances are identical for both spin directions.  
Once SOC is included the spectra depend on the spin direction indicating a magnetic ground state.
As the QMC temperature is above the anisotropy energy, ensemble averaging over both magnetic ground states is present so that the breaking of spin-symmetry is enhanced by including a small magnetic field.
On the flat surface, SOC shifts the resonance above $E_F$ and renders the resonance spin polarized, which is reminiscent of the amplitude asymmetry of the spin-resolved spectral function of Co in the KKR simulations.
In contrast, a Kondo resonance would split in energy in a magnetic field but retain identical amplitudes for both spin directions.
Neglecting SOC, a broad resonance is present below $E_F$ for the atom at the chain. 
Including SOC, the resonance disappears and steps are observed close to zero bias. 
At $T=58$~K no spin excitations would be experimentally resolved. 
From orbital projected spectral functions (\sm), the signatures close to zero bias can be attributed to orbital excitations, reflecting the many degrees of freedom of the multi-orbital QMC calculations. 
The \sm\ further presents the self energies $\Sigma(i\omega_n)$ from QMC for all four scenarios.
Without SOC, the self energies of the individual orbitals are spin-independent and evolve linearly at low Matsubara frequencies $i\omega_n$ as is the case for a Fermi liquid. 
Taking SOC into account, however, the self energies depend on the spin directions and Fermi liquid behavior is absent.
QMC therefore favors a spin-polarized ground state without Kondo screening when SOC is no longer neglected.

In summary, experimental $dI/dV$ spectra of Co atoms at monatomic Cu chains on Cu(111) show inelastic spin excitations.
This implies that magnetic anisotropy is relevant and SOC consequently must not be neglected in theoretical models of Co on Cu.
DFT calculations predict a different spin orientation and a reduced magnetic anisotropy compared to Co on flat Cu(111), which are consistent with the experimental evidence.
Two independent theoretical frameworks confirm that the experimental spectra originate from spin excitations rather than a Kondo effect.
It may be useful to extend the experimental and theoretical approaches presented here to other materials combinations that have been interpreted in terms of the Kondo effect.
\begin{acknowledgements}
We thank Juba Bouaziz for fruitful discussions.
A.M.M., F.S.M.G., S.B., and S.L. are supported by the European Research Council (ERC) under the European Union’s Horizon 2020 program (ERC-consolidator grant 681405 DYNASORE).
A.I.L. acknowledges the support by the ERC via Synergy grant 854843-FASTCORR.
We acknowledge computing time at the supercomputing centers of Forschungszentrum Jülich and RWTH Aachen. 
\end{acknowledgements}

\bibliography{citations}
\end{document}


\title{Supplemental Material:\\
Spin excitation of Co atoms at monatomic Cu chains}

\author{Neda Noei} \affiliation{\ieap}
\author{Roberto Mozara} \affiliation{\thh}
\author{Ana M. Montero} \affiliation{\jul}
\author{Sascha Brinker} \affiliation{\jul}
\author{Niklas Ide} \affiliation{\ieap}
\author{Filipe S. M. Guimar\~aes} \affiliation{\supercomp}
\author{Alexander I. Lichtenstein} \affiliation{\thh} 
\author{Richard Berndt} \affiliation{\ieap}
\author{Samir Lounis} \affiliation{\jul} \affiliation{\due}
\author{Alexander Weismann} \email{weismann@physik.uni-kiel.de} \affiliation{\ieap}

\maketitle

\section{Supplemental DFT Results}

\subsection{DFT Density of states}

\begin{figure}[!ht]
	\centering
		\includegraphics[width=0.7\linewidth]{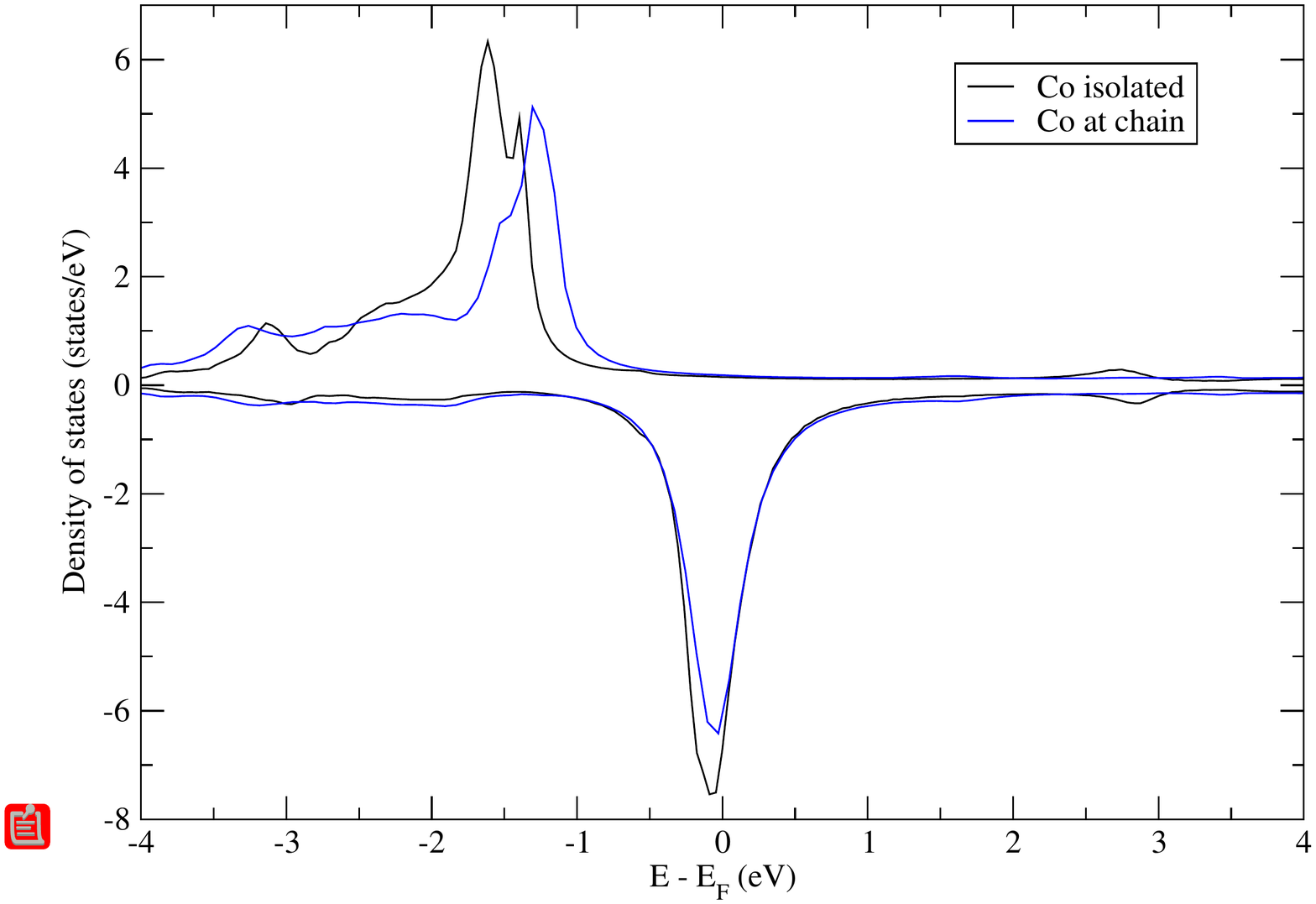}
		\caption{Spin-resolved $d$-orbital density of states of the Co atom on the surface (at a chain) calculated using ferromagnetic DFT with SOC.}
	\label{fig:MAE}
\end{figure}

\clearpage\newpage
\subsection{DFT magnetic anisotropy }
\begin{figure}[!ht]
	\centering
		\includegraphics[width=0.5\linewidth]{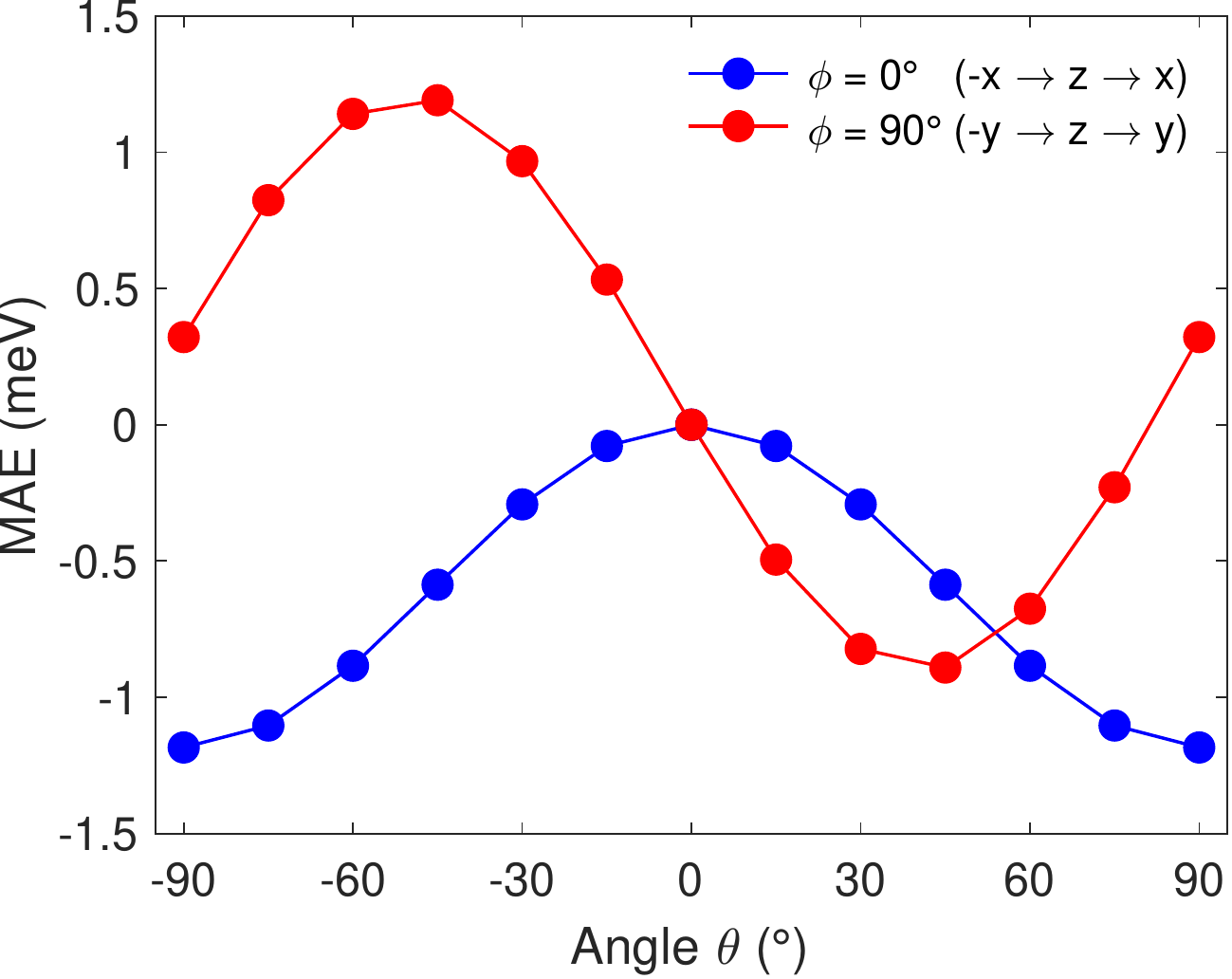}
		\caption{Magnetic anisotropy energy (MAE) of a Co atom at a chain derived from static DFT\@.
		$\theta$=0 is defined as the out-of-plane direction. 
		$\phi$=0 is oriented parallel to the chain. Large values correspond to preferred spin orientations.
		The hard axis is oriented along to the chain ($\phi=0$, $\theta=90^{\circ}$) and the easy axis is tilted by $\approx$ 45$^{\circ}$ from the out-of-plane direction towards the chain  ($\phi=90^{\circ}$, $\theta=-45^{\circ}$).
For $\phi=0^{\circ}$, $\theta$ runs from the $-x$-direction over $z$ to $x$. For $\phi=90^{\circ}$, $\theta$ sweeps from $-y$ over $z$ to $y$.}
	\label{fig:MAE2}
\end{figure}

\begin{figure}[!ht]
	\centering
		\includegraphics[width=0.5\linewidth]{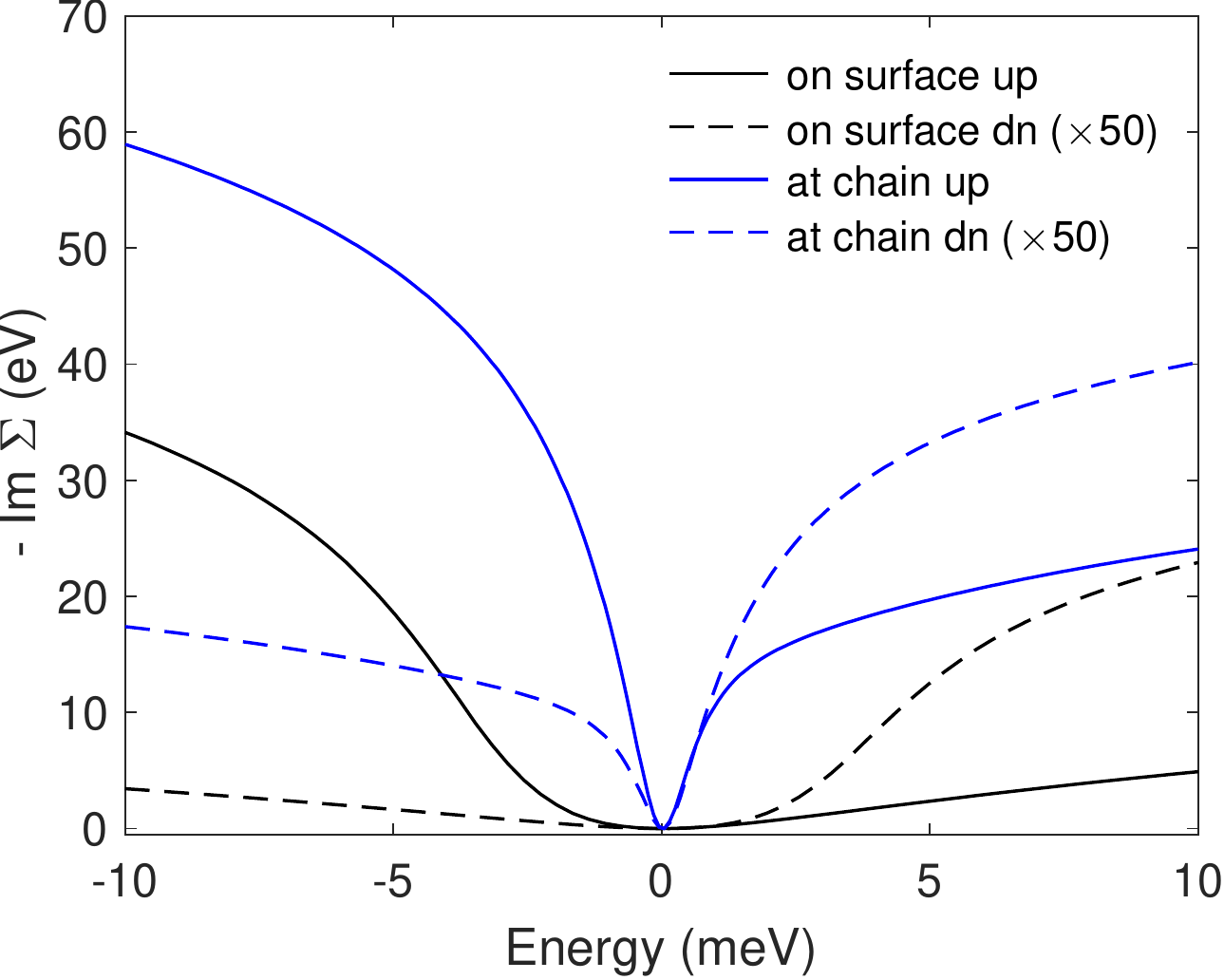}
		\caption{Imaginary part of the self energy within MBPT\@.
		Black (blue) lines show $-\mathrm{Im} \Sigma(\omega)$ for the Co atom on the surface (at the chain), while solid (dashed) lines represent majority (minority) spin directions.}
	\label{fig:Sigma_Spinaron}
\end{figure}
\clearpage

\newpage
\section{Supplemental CTQMC data}

\subsection{DFT and QMC setup}

The DFT input for the multi-orbital Anderson impurity model (AIM) was determined using the VASP code \cite{Kresse_PRB} with a $6\times6\times5$ slab for the Cu(111) surface, a vacuum separation of 12.5\,\AA, and a $9\times9\times1$ $\Gamma$-centered $\mathbf k$-point mesh.
Relaxation of the Co adatom and the surface was performed until the forces were below 0.02 eV/\AA\@.
The DFT bands within the energy window of $\pm$1.5 eV around the Fermi energy were projected onto the AIM with the method of projected local orbitals (PLO) \cite{Karolak_JPCM}.
A rotationally invariant Coulomb matrix with Slater parameters $U$\,$=$\,4.0 eV and $J$\,$=$\,0.9 eV was employed for both geometries, on the surface and at the chain. 
Crystal field splittings and hybridization functions, however, were different for both geometries.
The AIM was solved by the TRIQS/CTHYB code \cite{Parcollet_CPC, Seth_CPC}, and the QMC data were analytically continued using SOM \cite{Krivenko_GHub}.
The SOC strength was determined from the partial wave of the lowest bound valence state of the Co $d$-electrons and was consistent with the one reported in Ref.~\cite{dai_2008}.
To define the chemical potential, the ``fully localized limit'' (FLL)-approximation has been chosen, which resulted in a $S = 1$ scenario for both studied geometries. 
\subsection{SOC in QMC}
For consideration of the SOC in QMC we used two approximations: First, the PLO projectors were determined from paramagnetic DFT, so crystal field and hybridization function do not depend on the spin.
Second, the spin in QMC was assumed to be collinear because the local Hilbert space would otherwise be too large.
Including SOC initially produced a strong sign problem as eigenstates that initially belonged to different invariant subspaces of the local Hamiltonian were mixed. 
Transforming the local orbital basis from cubic into spherical harmonics turned out beneficial and resulted in a simpler expression for the Coulomb matrix.
Even though crystal field and hybridization acquired off-diagonal elements in the spherical basis, the sign problem was reduced by one order of magnitude, and the calculations became feasible.

\subsection{Spectral functions from CTQMC}

\begin{figure}[!ht]
	\centering
		\includegraphics[width=0.94\linewidth]{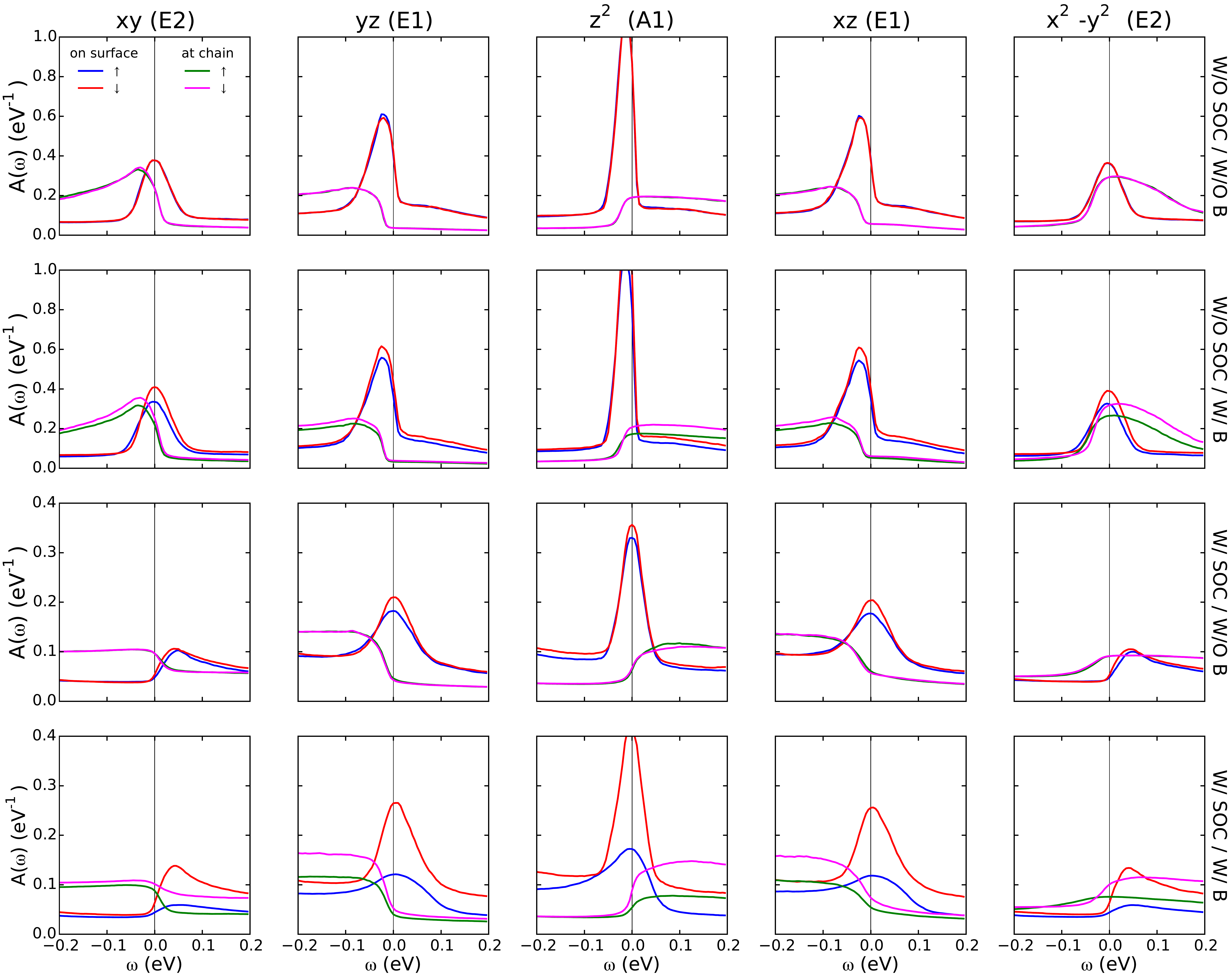}
		\caption{Orbital-resolved spectral functions $A(\omega)$ for different $d$-orbitals (columns) for the Co atom on the free surface and attached to the chain.
		Top row: Without SOC the spectra show Kondo resonances (simulation temperature $k_B T = 5$\,meV lies within the Kondo crossover scale \cite{surer_2012}). Orbitals of E2 symmetry (d$_{xy}$ and d$_{x^2-y^2}$), which show similar spectra on the free surface, differ in their spectral density when attached to the chain. 2nd row: A small magnetic field ($B_z=10$\,T) changes the spectra only slightly indicating its small scale ($\mu_B B_z \simeq 0.58$\,meV) and spin screening. 3rd row: With SOC ($\lambda \simeq 69.5$\,meV) the resonance positions are shifted to higher energies for the Co atom on the free surface while step-like spectral signatures are observed at the chain. Bottom row: With SOC, the spectra now react more strongly to the small magnetic field indicating symmetry breaking along the easy axis and a magnetic ground state.}
	\label{fig:QMC_orbspec_lowenergy}
\end{figure}

Without SOC and external magnetic field multi-orbital Kondo resonances emerge (1st row in Fig.~\ref{fig:QMC_orbspec_lowenergy}), cf. Ref. \cite{surer_2012}. 
The Kondo resonances of the Co adatom at the chain are broader and their positions are shifted compared to the Co atom on the free surface. 

A small magnetic field leads to no pronounced changes in the spectral behavior, while peaks near the Fermi energy can still be observed (2nd row in Fig.~\ref{fig:QMC_orbspec_lowenergy}). 
The magnetic field is an order of magnitude smaller than the Kondo scale, so that the Kondo effect prevails.

Including SOC leads to several spectral changes (3rd row in Fig.~\ref{fig:QMC_orbspec_lowenergy}). 
For the case of the Co atom on the free surface, the resonances are shifted to and above the Fermi energy. 
For the Co atom attached to the chain step-like features emerge and no clear peaks can be observed close to E$_F$.

\begin{figure}[!ht]
	\centering
		\includegraphics[width=0.9\linewidth]{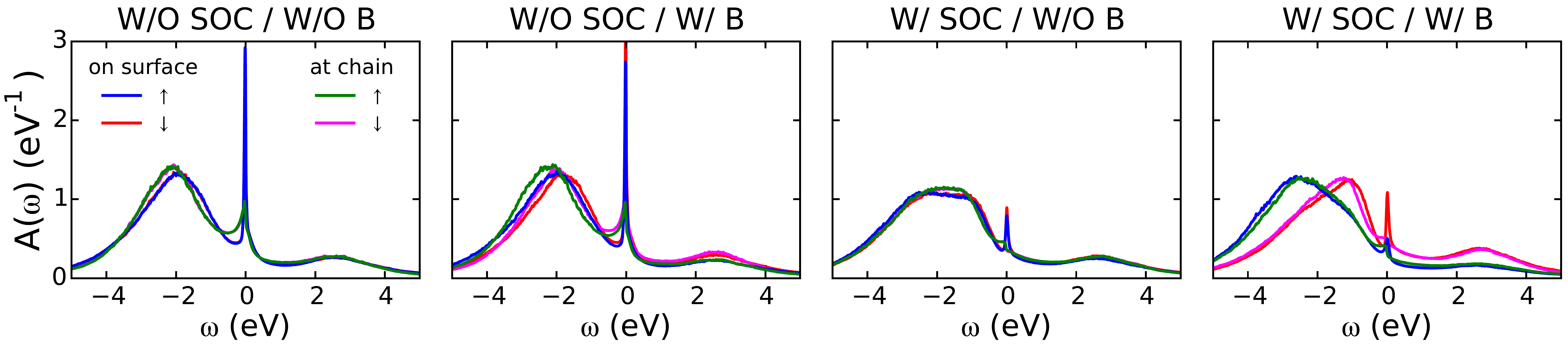}
		\caption{Orbital-integrated spin-resolved spectral functions $A(\omega)$ for the Co atom on the free surface and attached to the chain.
		1st and 2nd graphs: Without SOC, a small magnetic field  changes the spectra only slightly. The Kondo effect remains intact while spin-spectral weight is slightly shifted (spin-up is majority).
		3rd and 4th graphs: With SOC ($\lambda \simeq 69.5$\,meV) and a small magnetic field the spectrum becomes strongly spin-dependent indicating a magnetic ground state.
	\label{fig:QMC_spinspec_largeenergy}}
\end{figure}

SOC leaves the ground state two-fold degenerate with high $J_z$ expectation values (of opposite sign).
Transitions between the two remaining degenerate ground states in presence of SOC are only possible via higher-order hybridization events and thus are rather unlikely, so that a resulting magnetic moment exists. 
However, the simulation temperature is too high for the local moment to develop, {\em i.\,e.\@} for spontaneous symmetry breaking to occur.
A small magnetic field, however, may support the symmetry breaking so that the spectra become strongly spin-dependent when SOC is included (4th row in Fig.~\ref{fig:QMC_orbspec_lowenergy}). 
Orbital-integrated spin-resolved spectral functions are displayed in Fig.~\ref{fig:QMC_spinspec_largeenergy} on a larger energy scale. 

\begin{figure}[!ht]
	\centering
		\includegraphics[width=0.94\linewidth]{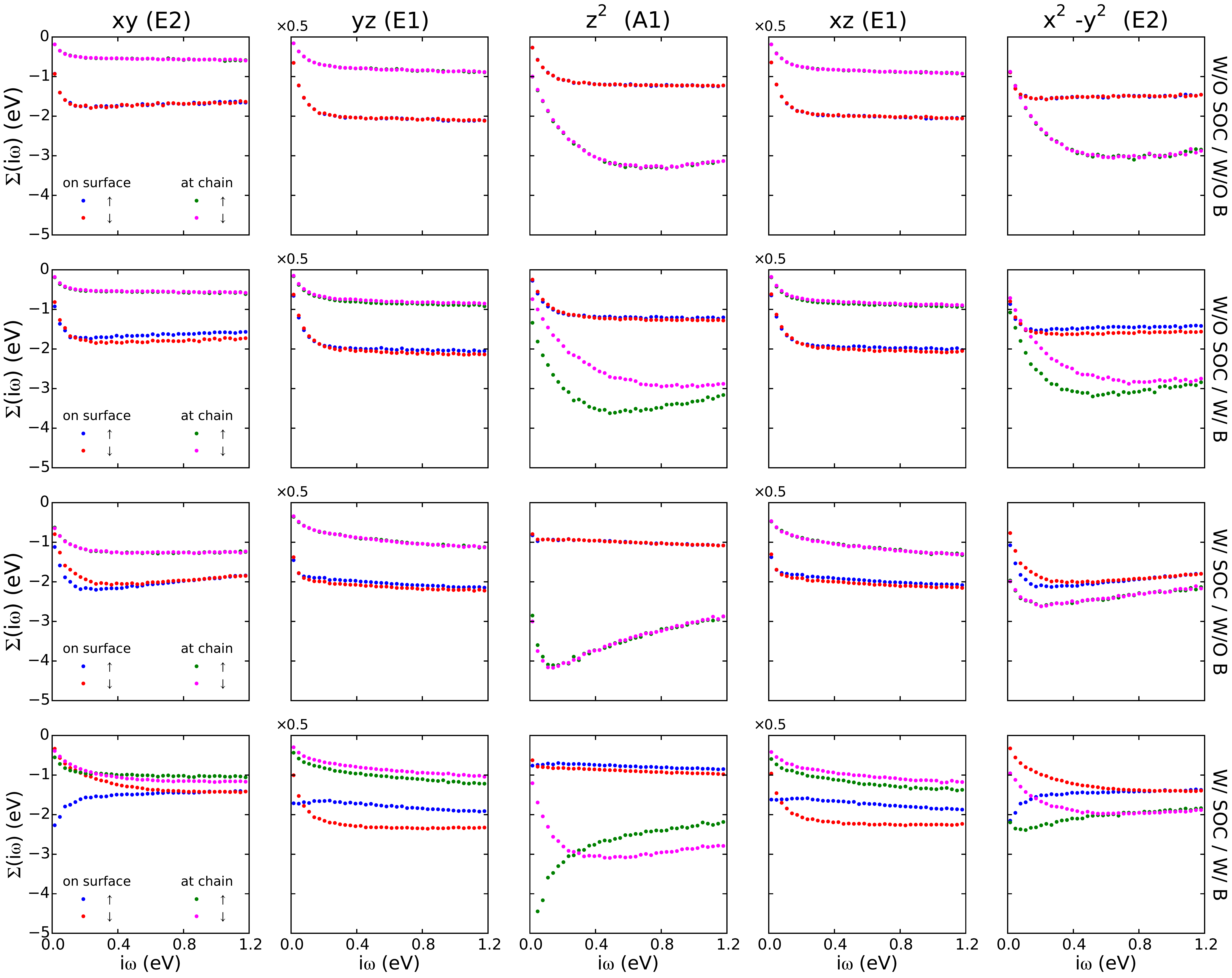}
		\caption{Imaginary part of the self-energy vs.\ Matsubara frequency $\Sigma(i\omega_n)$  for different $d$-orbitals (columns) for the Co atom on the free surface and at the chain.
		1st row: Without SOC the self-energies show Fermi-liquid (FL) behaviour as expected for Kondo systems. For lower Matsubara frequencies $-$Im$\Sigma$ tends to decrease linearly with a remaining scattering rate influenced by temperature. 2nd row: A small magnetic field ($B_z=10$\,T) changes the spectra only slightly. Spin screening and FL behavior is still present, however orbitals of $e_g$ symmetry ($d_{z^2}$ and $d_{x^2-y^2}$) of Co attached to the chain show reduced scattering rates for the minority spin channel. 
		3rd row: With SOC and in the presence of the easy axis FL behavior is reduced in orbitals of $e_g$ symmetry but a spin splitting of $\Sigma(i\omega_n)$ is not discernible because QMC performs an ensemble averaging over both magnetic ground states with the given input.
		4th row: A small magnetic field, however, can induce a significant spin splitting of $\Sigma(i\omega_n)$ of all orbitals. The presence of a local moment destroys FL behavior in some majority spin channels.
}
	\label{fig:Sigma_iw}
\end{figure}

Self-energies $\Sigma(i\omega_n)$ vs.\@ Matsubara frequency $i\omega_n$ are shown in Fig.~\ref{fig:Sigma_iw}.
When including SOC the states with less weight at the Fermi level exhibit non-Fermi liquid (NFL) behavior due to enhanced scattering rates. 
On the free surface, NFL behavior can be observed in all orbitals, while next to the chain it is prominent in the $d_{z^2}$ and to some extent in the $d_{x^2-y^2}$ orbital. 

\clearpage\newpage

\subsection{Self energies and non-Fermi liquid behavior}

\clearpage\newpage

\section{Supplemental line-shape analysis}
\begin{figure}[b]
	\centering
		\includegraphics[width=0.75\linewidth]{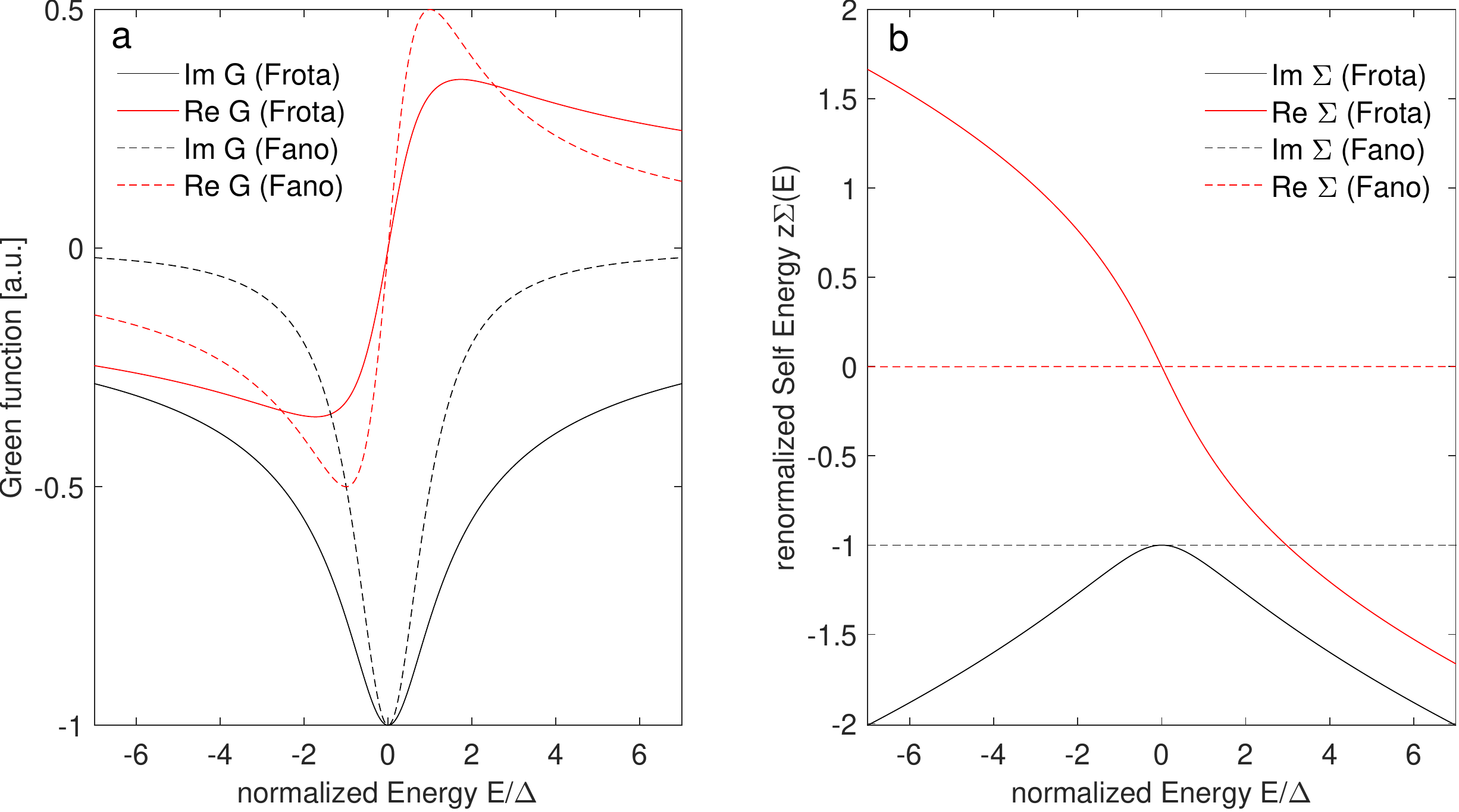}
		\caption{Comparison of a) Green functions and b) corresponding self energies of Fano and Frota resonances.
		A Fano resonance is related to a constant, energy independent broadening $\mathrm{Im} \Sigma(E)=-\Delta$. 
		A Frota resonance in contrast has a self-energy which imaginary (real) part shows to lowest order quadratic (linear) behavior with $E - E_K$, as it is the case for a Fermi liquid. The overall shape shows remarkable similarities to the self energy of Kondo-systems calculated by advanced many-body methods (compare e.\,g.\ Figure 3 in \cite{bulla_1998}) }
	\label{fig:Sigma_FanoFrota}
\end{figure}

As briefly mentioned in the manuscript, lifetime-broadened single-particle states, Kondo resonances and inelastic excitations have distinctly different spectroscopic line-shapes. 
In the first case, the Green function is a complex Lorentzian with spectral weight $z$ and width $\Delta$ that is centered around $E_K$:
\begin{equation}
G_L(E)=\frac{z}{E-E_K+i\Delta}
\end{equation}
The real (imaginary) part of $G_L(E)$ is shown in Fig.~\ref{fig:Sigma_FanoFrota} by a red (black) corresponds to a Fano resonance \cite{fano_1961} with line shape parameter $q$ = 1 ($q$ = 0).
Furthermore, all possible Fano asymmetries ({\em i.\,e.\@} $q \in \left[-\infty,\infty\right]$) may be obtained from $G_L(E)$ using a phase $\phi_q=2\cdot\mathrm{tan}^{-1}(q)$ via $dI/dV(V)~\propto \mathrm{Im}\left[ e^{i\phi_q}G_L(E)\right]$.

In contrast to Fano line-shapes, Kondo resonances show an inverse square root behavior, a smaller phase shift  between $E << E_K$ and $E >> E_K$ \cite{pruser_2011,pruser_2012} and are well described by the Green function derived by Frota \cite{frota_1992}:
\begin{equation}
G_F(E)=-iz\sqrt{\frac{i}{E-E_K+i\Delta}}
\end{equation}
Real (imaginary) part of $G_F(E)$ are shown in Fig.~\ref{fig:Sigma_FanoFrota} by red (black) solid lines. 
The differences between both expressions can be emphasized by analyzing the self-energy $\Sigma(E)$ that may be obtained using:
\begin{equation}
\Sigma_{L,F}(E)=(G_{L,F}(E)^{-1}-E)^{-1}
\end{equation}

\begin{figure}[!h]
	\centering
		\includegraphics[width=0.6\linewidth]{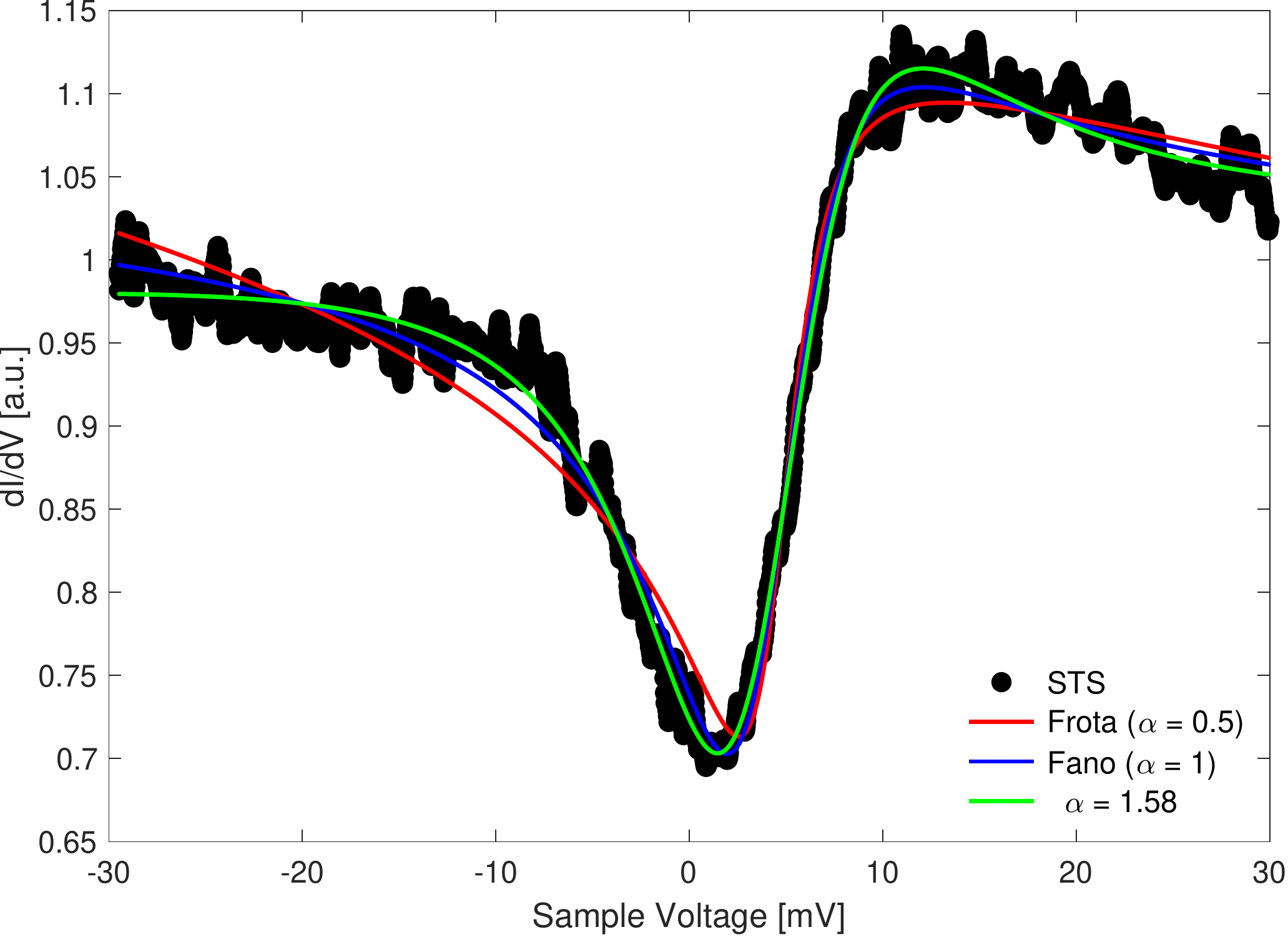}
		\caption{Experimental $dI/dV(V)$ spectrum recorded on a single Co atom on the free surface (black dots) along with Fits using different expressions.
		The Frota expression (red curve) shows poor agreement while a fit using the Fano expression (blue curve) is superior. When $\alpha$ is left as an adjustable parameter (green curve) a line-shape "beyond" the Fano resonance ($\alpha>1$) is favored.}
	\label{fig:alpha_EXP}
\end{figure}

The real (imaginary) parts of $\Sigma(E)$ correspond to a shift (broadening) of spectral signatures. 
It can be seen from Fig.~\ref{fig:Sigma_FanoFrota}b that Fano resonances result from an energy-independent self-energy, as it is the case when a discrete state is hybridized with a flat, energy-independent density of states. 
In contrast, the self energy of a Frota resonance is energy-dependent and its imaginary part (real part) is to lowest order quadratic (linear) with $E - E_K$.
This behavior is characteristic of a Fermi liquid. 
Although the above expression for $G_F(E)$ is of phenomenological nature, the resulting self-energy shown in Fig.~\ref{fig:Sigma_FanoFrota}b is in remarkable agreement with $\Sigma(E)$ of Kondo systems calculated using advanced many-body methods (compare e.g. with Figure 3 in \cite{bulla_1998}).

To determine which of both line-shapes agrees better with measured and calculated spectra we introduce the following expression for fitting Fano and Frota line-shapes using the parameter $\alpha$:
\begin{equation}
\frac{dI}{dV}(V)\propto \mathrm{Im}\left[-ie^{i\phi_q}\left(\frac{i}{\epsilon-E_K+i\Delta}\right)^{\alpha}\right]
\end{equation}
Here $\alpha=0.5$ ($\alpha=1$) corresponds to a Frota (Fano) resonance. The phase $\phi_q$ defines the line-shape asymmetry as described above and is related to the $q$-parameter defined in \cite{fano_1961} by $q=tan(\phi_q/2)$. 
We leave $\alpha$ as an adjustable parameter to determine whether a specific resonance is closer to the Fano or Frota line shape.

\begin{figure}[!h]
	\centering
		\includegraphics[width=0.6\linewidth]{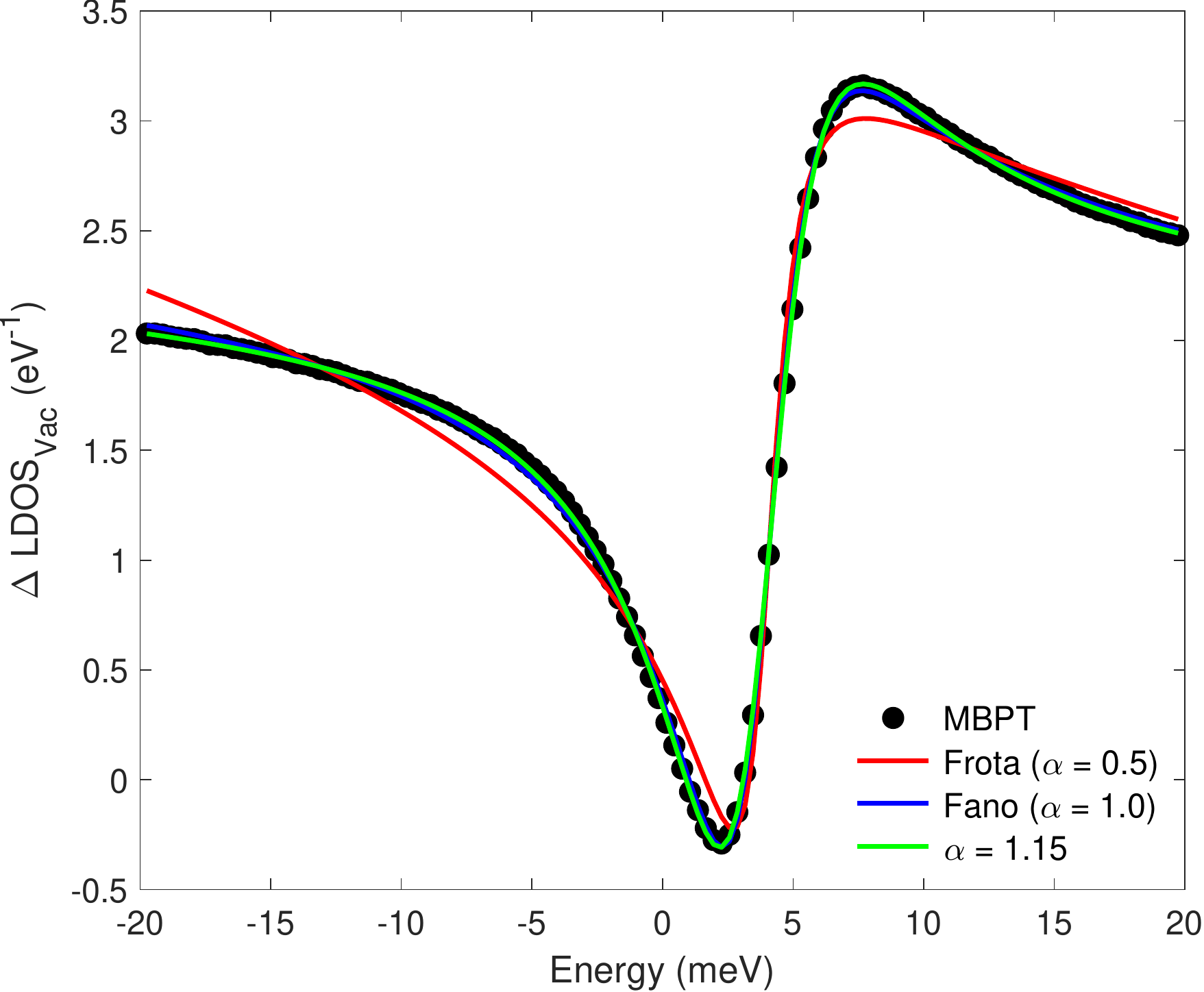}
		\caption{Vacuum LDOS calculated by DFT+MBPT (black dots) showing a spinaron at positive energies and fits using different expressions.
		The Fano resonance (blue curve) shows better agreement than the Frota lineshape (red curve).
		When $\alpha$ is treated as an adjustable parameter, a resonance close to a Fano shape is favored (green curve).}
	\label{fig:alpha_THEO}
\end{figure}

Figure~\ref{fig:alpha_EXP} shows an experimental $dI/dV$ spectrum recorded on a single Co atom on the pristine surface.
The Fano resonances provide a better fit than Frota lines, albeit not being perfect.
Leaving $\alpha$ as a free parameter, the optimization favors $\alpha>1$ demonstrating that the measured spectrum does not exhibit the characteristic shape of a Kondo resonance.

Figure ~\ref{fig:alpha_THEO} displays an identical analysis of a spectrum calculated within DFT+MBPT\@.
In this case, the best fit is observed for $\alpha$ close to one.
In other words, a Fano resonance describes the Spinaron signature well.
We conclude that the Spinaron line shape is distinctly different from that of a Kondo resonance.  

\bibliography{citations}